\documentclass[%
 aip,
 apl,%
 amsmath,amssymb,
 reprint,%
]{revtex4-1}

\usepackage{color}
\usepackage{graphicx}
\usepackage{latexsym}
\usepackage{textcomp}

\begin{document}

\title {Electrical transport in C-doped GaAs nanowires: surface effects}

\author{Alberto Casadei} 
\affiliation{ Laboratory of Semiconductor Materials, Ecole Polytechnique F\'{e}d\'{e}rale de Lausanne, CH-1015 Lausanne, Switzerland}

\author{Jil Schwender} 
\affiliation{Institute of Applied Physics, University of Hamburg, Jungiusstr. 11, D-20355 Hamburg, Germany}

\author{Eleonora Russo-Averchi} 
\affiliation{ Laboratory of Semiconductor Materials, Ecole Polytechnique F\'{e}d\'{e}rale de Lausanne, CH-1015 Lausanne, Switzerland}

\author{Daniel R\"{u}ffer} 
\affiliation{ Laboratory of Semiconductor Materials, Ecole Polytechnique F\'{e}d\'{e}rale de Lausanne, CH-1015 Lausanne, Switzerland}

\author{Martin Heiss} 
\affiliation{ Laboratory of Semiconductor Materials, Ecole Polytechnique F\'{e}d\'{e}rale de Lausanne, CH-1015 Lausanne, Switzerland}

\author{Esther Alarc\'o- Llad\'o}
\affiliation{ Laboratory of Semiconductor Materials, Ecole Polytechnique F\'{e}d\'{e}rale de Lausanne, CH-1015 Lausanne, Switzerland}

\author{Fauzia Jabeen} 
\affiliation{ Laboratory of Quantum Optoelectronics, Ecole Polytechnique F\'{e}d\'{e}rale de Lausanne, CH-1015 Lausanne, Switzerland}

\author{Mohammad Ramezani} 
\affiliation{ Laboratory of Semiconductor Materials, Ecole Polytechnique F\'{e}d\'{e}rale de Lausanne, CH-1015 Lausanne, Switzerland}

\author{Kornelius Nielsch}
\affiliation{Institute of Applied Physics, University of Hamburg, Jungiusstr. 11, D-20355 Hamburg, Germany}

\author{Anna Fontcuberta i Morral}
\affiliation{ Laboratory of Semiconductor Materials, Ecole Polytechnique F\'{e}d\'{e}rale de Lausanne, CH-1015 Lausanne, Switzerland}

\date{\today}

\begin{abstract}
The resistivity and the mobility of Carbon doped GaAs nanowires have been studied for different doping concentrations. Surface effects have been evaluated by comparing upassivated with passivated nanowires. We directly see the influence of the surface: the pinning of the Fermi level and the consequent existence of a depletion region lead to an increase of the mobility up to 30 cm$^2/($V$\times$s$)$ for doping concentrations lower than $3\times 10^{18}$ cm$^{-3}$. Electron beam induced current measurements show that the minority carrier diffusion path can be as high as $190$ nm for passivated nanowires.
\end{abstract}

\maketitle

\section{Introduction}
The particular morphology and dimensions of nanowires (NWs) have opened in the last years many perspectives in applications ranging from electronics, optoelectronics to energy harvesting and storage\cite{Tom12,Wal13,Kro13}. The successful incorporation of dopants in the nanowires is key for the implementation of such devices. Doping of nanowires has been a challenging task, due to the particular growth mechanisms\cite{Wal11}. 
In this work we report on the electrical properties of C-doped GaAs nanowires obtained by the Ga-assisted method\cite{Col08}. As C is not soluble in Ga, its incorporation in the nanowire core through the droplet remains difficult. Our strategy is then to grow an intrinsic Nw core and to grow a doped shell around it. We study the doping and surface effects in NWs resistivity, field effect mobility as well as the minority carrier diffusion length. 

\section{Method}
The nanowires consisted of an undoped GaAs core of about 70 nm in diameter and a C-doped shell. The GaAs nanowires were grown on a (111) Si substrate under conditions that maximize the yield of vertical nanowires, as described elsewhere\cite{Ele12,Ucc11}: a rotation of 7 rpm, a flux of Ga equivalent to a planar growth rate of $0.3$ $\textrm{\AA/s}$, a substrate temperature of $640$ \textcelsius$ $ and a V/III beam equivalent pressure (BEP) ratio of 50. Subsequently to the GaAs nanowire synthesis, growth conditions were radically changed for the obtaining of a $40$ nm GaAs shell. The Ga flux was stopped for about 10 min while the substrate temperature was lowered to $465$ \textcelsius$ $ and the V/III BEP ratio was increased to $130$\cite{Hei09}. A p-type doping was achieved by adding a flux of Carbon during the growth of the shell. To test the effect of C doping, we grew a series of samples with different C flux given by the current applied to the cell, as reported in table 1. Equivalent samples have been fabricated with the addition of a capping layer consisting of 15 nm Al$_{0.3}$Ga$_{0.7}$As and 6 nm of intrinsic GaAs. \\\indent
To give an accurate estimation the resistivity of the nanowires, multiple contacts were fabricated\cite{Duf10}. Multiple four point measurements have been performed on around 100 samples. The realization of a large number of devices has been automatized thanks to our auto-contacting software \cite{Bla13}. The electrical contacts consisted of Pd/Ti/Au ($40/10/250$ nm) metal layer directly in contact with the doped shell. For the capped nanowires, the Al$_{0.3}$Ga$_{0.7}$As shell was removed using a solution of citric acid and hydrogen peroxide (2:1) leading an etching rate of $1.4$ $\textrm{nm/s}$. Bottom gate transistors, with 200 nm thick SiO$_2$ layer used as a gate dielectric, were realized in order to extract the nanowire majority carrier mobility.

\section{Results and discussions}
Surface states play an extremely important role in the opto-electronic properties of GaAs nanowires. In ambient conditions there is a formation of a thin oxide layer on the NW surface resulting in a pinning of the Fermi level. In GaAs the pinning occurs near the middle of the band gap, producing a depletion region. The existence of this region reduces the electrically active part of the NWs\cite{Hei11}. We can avoid this problem by passivating the GaAs with a wider band gap semiconductor, such as AlGaAs\cite{Dem10,Tat04}. In this study we study how a passivation layer affects the electronic properties of GaAs NWs with different doping concentrations.\\\indent
7 types of samples have been analyzed and the shell doping concentration, the resistance and the mobility are reported in table 1. 
\begin{table}[htbp]
\centering
\begin{tabular}{lccc}
\hline
Sample & Doping conc. & Res/Length & Mobility \\
& [cm$^{-3}$] & [k$\Omega$/$\mu$m] & [cm$^2$/(V$\times$s)]\\
\hline
\hline
1c & $8\times 10^{18}$ & 91& 8\\

2c & $5\times 10^{18}$& 53 & 12\\

3c & $2.5\times 10^{18}$& 103 & 14\\

4c & $1\times 10^{18}$& 2700 & 20\\

1u & $4\times 10^{18}$& 329 & 8\\

2u & $2\times 10^{18}$& 300 & 20\\

3u & $1.3\times 10^{18}$ & 528 & 30\\

\hline
\end{tabular}
\label{tab1}
\caption{Doping shell concentration, resistance per unit length and mobility of the nanowires analyzed. "p" corresponds to the passivated NWs and "u" to the unpassivated ones.}
\end{table}

\begin{figure}[ht]
\begin{center}
\includegraphics[width=6 cm]{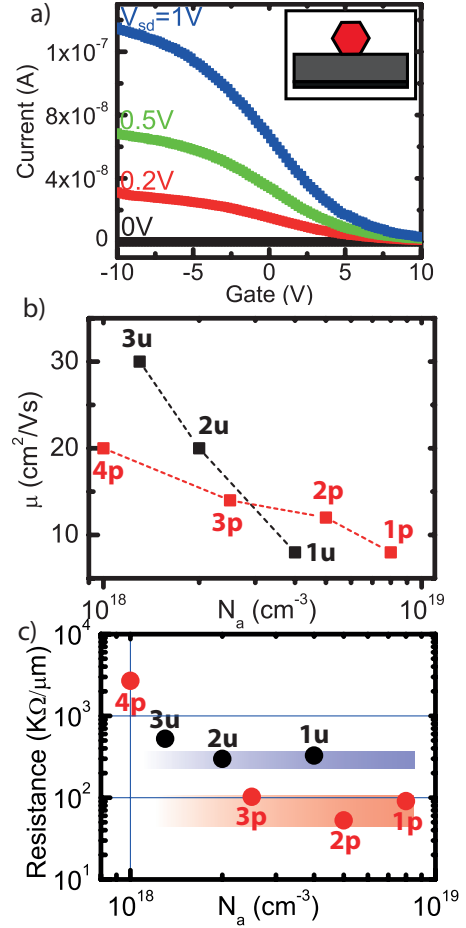}
\end{center}
\caption{a) Example of a transistor output curve for different source-drain voltages $V_{sd}$. The inset shows the NW bottom-gate transistor configuration. b) Mobility vs shell doping concentration for the 7 types of NWs analyzed. 1u, 2u and 3u are unpassivated NWs while 1p, 2p, 3p and 4p are passivated. c) NW resistance per length in function of the doping concentration. The black and red stripes highlight the region of minimum resistance per length for unpassivated and passivated NWs.}
\label{fig1}
\end{figure}

The conductivity $\sigma$ is obtained from 4 point measurements and the mobility $\mu$ is derived from bottom gate field effect transistor characteristics:
\begin{equation}
\mu= -\frac{\partial I}{\partial V_g} \frac{L^2}{C\times V_{sd}}
\end{equation}
where $V_g$ and $V_{sd}$ are respectively the gate and the source-drain voltages, $L$ is the distance between source and drain and $C$ is the device capacitance. For a structure formed by a nanowire gated through the substrate, it is found\cite{Kha07} :
\begin{equation}
C= 2\pi \epsilon_0 \epsilon_r \frac{L}{x\times cosh^{-1}\left(\left(r_0+h\right)/r_0\right)}
\end{equation}
where $\epsilon_0\epsilon_r$ is the dielectric constant of insulating layer, $r_0$ is the NW radius and $h$ the dielectric thickness. In the case of passivated NWs we consider the capping layer as a capacitance in series with the bottom gate dielectric. For 200 nm SiO$_2$ thickness with $\epsilon_r=3.9$ the typical capacitance value of our devices is around $1$ pF. Typical output curves of our devices are shown in Fig. \ref{fig1}a.\\\indent

We assume that all acceptors are ionized at room temperature, meaning that the holes concentration can be directly linked to the C concentration. As there is no depletion region within the GaAs doped shell of the passivated NWs (1c, 2c, 3c and 4c), we can consider that the electrically active part corresponds to the entire doped shell. We can therefore introduce the conductivity of the doped shell $\sigma_{shell}$ calculated on the doped cross section area and directly extract the doping density
\begin{equation}
N_{A(passivated)}=\frac{\sigma_{shell}}{\mu\times e}
\end{equation}

When NWs are not passivated, one needs to consider the Fermi level pinning at the NW surface and the existence of a depletion layer. Solving the Poisson equation and using the boundary conditions of a vanishing electric field at the surface we can calculate the depletion region width. For the doping concentration analyzed, the surface potential $\Phi$ at the GaAs surface is found to be around $0.5$ eV\cite{Lan84}. The depletion region $w$ follows\cite{Cas13}:
\begin{equation}
-\frac{w^2}{2}+r_0 w -(r_0-w)^2 ln \left( \frac{r_0}{r_0-w} \right)=\frac{2\epsilon_r \epsilon_0 \Phi}{e N_{shell}}
\end{equation}
where $\epsilon_r=\epsilon_{GaAs}=12.9$ is the GaAs dielectric constant and $N_{shell}$ the doping concentration in the shell. $N_{shell}$ and $w$ can be deduced from the electrical measurements using the following system:
\begin{equation}
\left\{
\begin{array}{l}
\int_{r_{core}} ^{r_{0}-w} \int_0 ^{2\pi} N_{shell}x sin(\theta) dx d\theta = \frac{L}{\mu R e} \\
-\frac{w^2}{2} + r_0 w - (r_0 - w)^2 ln\left( \frac{r_0}{r_0 - w}\right) - \frac{2 \epsilon_r \epsilon_0 \Phi} {e N_{shell}} = 0 
\end{array}
\right.
\end{equation}
where $R/L$ is the resistance per unit length, $\mu$ is the carrier mobility and $r_{core}$ is the radius of the undoped nanowire core.\\\indent

\begin{figure}[ht]
\begin{center}
\includegraphics[width=\columnwidth]{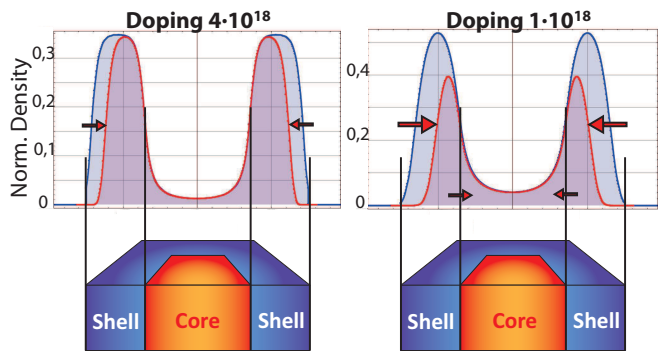}
\end{center}
\caption{Transverse section view of 2 NWs with intrinsic core and a doped shell ($4\times 10^{18}$ and $1\times 10^{18}$ cm$^{-3}$). The distribution holes profile is represented in the case of NWs capp (blue lines) and NWs uncap (red lines). In the unpassivated NWs, the depletion region has the effect of keeping the holes far from the surface. The numerical analysis has been performed with $nextnano$.}
\label{fig2}
\end{figure}

Results on the mobility and resistance per length obtained for all different types of nanowries are shown in Fig \ref{fig1}b and \ref{fig1}c. The mobility measured increases inversely with the doping concentration for both passivated and unpassivated NWs. This behavior is commonly observed in semiconductors and is related to the scattering of carriers by the doping impurities. Interestingly, we observe a reduction of mobility with the increase in carrier concentration, which results in the saturation of the resistivity for doping concentrations higher than $2\times 10^{18}$ cm$^{-3}$ . The saturation values correspond to $\approx 80$ k$\Omega$/$\mu$m and $\approx 330$ k$\Omega$/$\mu$m respectively for the passivated and unpassivated NWs. \\\indent
A priori one would expect a higher mobility for passivated nanowires with respect to the unpassivated. In Fig. \ref{fig1}b we observe that this is only true for the doping concentrations higher than $3\times 10^{18}$ cm$^{-3}$. In order to shed some light to this issue, we have simulated the band profile across the nanowire with $nextnano^3$ software  (copyright (c) 2012, nextnano GmbH, Germany) and calculated the carrier distribution inside the NWs. Figure 2 shows the profile of the hole concentration for both passivated and unpassivated NWs and for two doping concentrations: $4\times 10^{18}$ cm$^{-3}$ and $1\times 10^{18}$ cm$^{-3}$ (Fig. \ref{fig2}).  In the unpassivated case the Fermi level pinning was used.

\begin{figure}[ht]
\begin{center}
\includegraphics[width=\columnwidth]{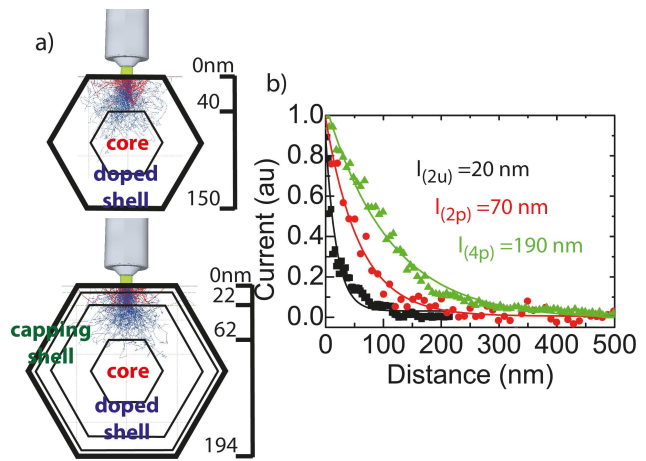}
\end{center}
\caption{a) Electron trajectory in the NWs simulated by Monte Carlo. The simulations are performed for 200 electrons with energy of $3$ keV and a probe current of $300$ pA. b) Experimental current decay as a function of distance to the contact for 3 NWs samples (2u, 2p and 4p). The minority carrier mean free path extracted increase up to $190$ nm for passivated NWs with a doping concentration of $10^{18}$ cm$^{-3}$.}
\label{fig3}
\end{figure}

As shown in Fig. \ref{fig2} holes are mainly distributed in the doped shell, but a small portion is also found in the core. In the case of unpassivated wires, the surface depletion leads to a decrease in the amount of carriers in the shell. This is especially notable for the low doping concentration. From this we conclude that two mechanisms may contribute to the observation that the highest mobility is for low doped unpassivated NWs. First, the depletion region in unpassivated and low doped NWs ensures that transport of carriers occurs at a certain distance from the surface and reduces the surface-related scattering. Second, a higher proportion of the transport occurs through the undoped core, which exhibits highest mobility due to the absence of dopants. \\\indent
Finally, we used electron beam induced current (EBIC) to study the mean free path of minority carriers. The semiconductor-metal interface at the contacts produces a Schottky barrier. When exciting electron-hole pairs with light or an electron beam, the potential gradient produced by the Schottky barrier attracts minority carriers (electrons) into the contacts, which is measured by an electrical current. The current exponentially decays with the distance from the contact \cite{All09}:
\begin{equation}
I=I_0 e^{x/l}+c
\end{equation}
where $l$ is the electron mean free path. In order to understand what region of the nanowire is excited with the electron beam, we simulate the trajectory of electrons from the beam with Casino \cite{Dro07}, shown in (Fig. \ref{fig3}a). We simulate the trajectories of 200 incident electrons with an energy of $3$ keV and a probe current of $200$ pA, corresponding to the experimental conditions used. As it can be seen in Fig. \ref{fig3}a, the excitation is limited to the doped shell of the nanowire.

Figure \ref{fig3}b shows the exponential decay of the current for an unpassivated and two passivated NWs with hole concentrations respectively of $2,4$ and $1\times 10^{18}$ cm$^{3}$. The lowest mean free path is $20$ nm for the unpassivated nanowire. The passivated nanowires exhibit longer mean free paths for minority carriers, up to $190$ nm. Despite the majority carrier (holes) mobility is lower for passivated NWs, a longer diffusion path is found for the minority carriers (electrons). This is explained again by the Fermi level pinning at the surface and its respective electric field, which attracts the beam-induced electrons towards the surface. Consequently, the electron main free path in unpassivated NWs is reduced by the larger defect and trap densities at the surface.

\section{Conclusions}
In conclusion, we have measured the carrier concentration, mobility and minority diffusion length for core-shell GaAs nanowires were the shell is C-doped at different concentrations. The effect of the surface states is evaluated by comparing the values for passivated and unpassivated NWs. At low doping concentrations the highest mobility is observed for unpassivated nanowries, while at higher doping concentrations the passivated nanowires offer the best characteristics. The highest mobility (30 cm$^2$/(V$\times$s)) was obtained for unpassivated NWs with a doping concentration of $1.3 \times 10^{18}$ cm$^2$/(V$\times$s). Finally, we have measured the minority carrier diffusion lengths up to $190$ nm for passivated NWs with a doping concentration of $10^{18}$ cm$^2$/(V$\times$s).

\section*{Acknowledgement}
The authors thank financial support from: The Swiss National Science Foundation under Grant No. 2000021-
121758/1 and 129775/1;
NCCR QSIT; The European Research Council under Grant "Upcon".

\end{document}